\documentclass[prl,twocolumn]{revtex4}

\usepackage{graphicx}
\usepackage{amsmath}

\begin{document}

\title{Controlled Creation of Spatial Superposition States for Single
  Atoms}

\author{Kieran Deasy$^{1,2}$, Thomas Busch$^{3,4}$, Yueping Niu$^5$,
  Shangqing Gong$^5$, Shiqi Jin$^5$ and S\'ile Nic Chormaic$^{1,2,3}$}

\date{\today}

\begin{abstract}
  We present a method for the controlled and robust generation of
  spatial superposition states of single atoms in micro-traps. Using a
  counter-intuitive positioning sequence for the individual potentials
  and appropriately chosen trapping frequencies, we show that it is
  possible to selectively create two different orthogonal
  superposition states, which can in turn be used for quantum
  information purposes.
\end{abstract}

\affiliation{$^{1}$Dept.~of Applied Physics and Instrumentation, Cork
  Institute of Technology, Bishopstown, Cork, Ireland\\
  $^{2}$Tyndall National Institute, Prospect Row, Cork, Ireland\\
  $^{3}$Physics Department, University College Cork, Cork, Ireland\\
  $^{4}$Physics Department, Dublin Institute of Technology, Kevin St.,
  Dublin 2, Ireland\\
  $^{5}$Key Laboratory for High Intensity Optics, Shanghai Institute
  of Optics and Fine Mechanics, Shanghai 201800, P.R. China}
\maketitle

During recent years trapping and controlling small numbers of neutral
atoms has emerged as one of the most active and productive frontier
areas in research
\cite{Haensel:01,Bergamini:04,Chuu:05,Shevchenko:06}.  The interest in
single atom systems is driven not only by the desire to perform
experiments that answer longstanding and fundamental quantum mechanics
questions \cite{Eschner:01,Beugnon:06}, but also by the desire to
implement concepts of quantum information using neutral atoms
\cite{Brennen:99,Jaksch:99,Jaksch:00}.  Advances in the technology of
optical lattices and micro-traps have recently allowed for substantial
progress in this area
\cite{Birkl:01,Rauschenbeutel:05,Yavuz:06,Sortais:06} and various
concepts have been developed to prepare and process the states of
single atoms in a controlled way. While techniques for controlling and
preparing the internal states of atoms using appropriate
electromagnetic fields are well developed, only few concepts exist for
achieving the same control over the spatial part of a wavefunction
\cite{Zhang:06,Eckert:04, Eckert:06}.

Three fundamental requirements for controlling the spatial part of a
wavefunction are preparation, storage and transport. Optical and
magnetic micro-potentials have proven to be robust tools for atom
storage and techniques for moving particles between different
potential sites using dedicated waveguides or sophisticated tunneling
schemes have been suggested \cite{Arlt:01,Brugger:05, Mompart:03}.
While waveguides are usually highly static, the tunneling interaction
can be tuned by actively changing the distance between or the
potential heights of neighboring traps.  Both of these possibilities
have recently been explored by Eckert {\sl et al.}  \cite{Eckert:04,
  Eckert:06} and Greentree {\sl et al.}  \cite{Greentree:04}. These
works considered three modes in three separated potentials and
suggested the use of a STIRAP-like process to achieve a robust
transfer of an atom from one trap to another with high fidelity. In
the area of three-level optics, the STIRAP process refers to the
technique whereby a counter-intuitive application of laser pulses
leads to a transition of an electron between the ground states in a
$\Lambda$-system \cite{Bergmann:98,Vitanov:01}. In the atom trap
scenario the energy levels are replaced by spatially separated trap
ground states and the laser interaction is replaced by the coherent
tunneling interaction. Eckert {\sl et al.}  also showed analogues for
coherent population trapping and electromagnetically induced
transparency \cite{Eckert:04}. They termed this new area {\sl
  three-level atom optics}.

\begin{figure}[tb]
  \includegraphics[width=\linewidth,clip=true]{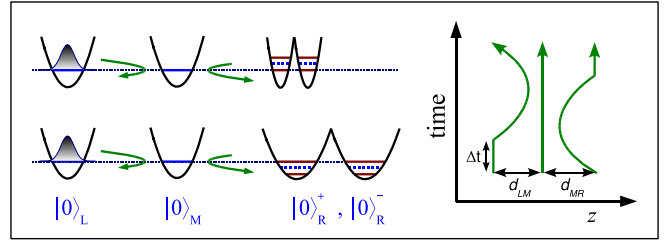}
  \caption{Left: The three atom traps are arranged in a linear
    fashion. The trap on the left and the trap in the middle are
    harmonic potentials and the double well trap on the right is made
    piecewise out of two harmonic traps. The trapping frequencies are
    arranged in such a manner that either the symmetric state (upper
    schematic) or the anti-symmetric state (lower schematic) of the
    double well is in resonance with the ground states of the other
    traps. The asymptotic ground states in the double well traps are
    indicated by a broken line.  Right: In the counter-intuitive
    timing scheme the distance between the double well and the middle
    trap, $d_{MR}$, is reduced before the distance between the middle
    trap and the trap on the left, $d_{LM}$.}
\label{fig:Schematic}
\end{figure}

In the following we will describe a process that extends the existing
work in three-level atom optics to the domain of four levels. In
particular we will show how to create a coherent superposition of an
atomic center-of-mass state in a controlled way and study its
stability. While this work is based on a recent suggestion for
multi-level optical systems \cite{Niu:04}, its translation into the
atomic realm does not only require considering new constraints on
time-scales, but also allows for an easy interpretation of the physics
involved.

The scheme is based on two well-known, quantum mechanical phenomena.
The first one is the above mentioned existence of dark states in
multi-level systems with an appropriately shaped coherent coupling. In
spatial arrangements such levels correspond to trap eigenstates and
the coupling is facilitated by a tunneling interaction. No direct
coupling is allowed between the initially occupied level and the final
state and this can be achieved by choosing a linear trap arrangement
(see Fig.~\ref{fig:Schematic}). A STIRAP-like process now requires the
tunneling interactions between neighboring potentials to increase and
decrease in accordance with a counter-intuitive timing sequence: first
the tunneling probability between the two empty states is increased
and after a specific delay, $\Delta t$, the tunneling interaction
between the occupied and empty state is increased. This leads to a
very robust transfer of the particle between the two states that
construct the dark state and in particular it avoids stringent
conditions on the timing of the interactions. In a two-level setting
such conditions are essential to avoid Rabi oscillations. The STIRAP
process has already been extensively analyzed in three-level optics
\cite{Bergmann:98,Vitanov:01}.
 
The second phenomenon is the appearance of ground state splitting in
double well potentials. When combining two single traps to form a
double well potential their respective asymptotic ground states
combine to yield a symmetric and an antisymmetric state, with the
energy difference between these two states depending on the distance
between the two traps. This energy difference is often referred to as
the tunneling splitting energy and is directly related to the
Rabi-frequency of the tunneling oscillations between the two wells.

Our setup is shown schematically in Fig.~\ref{fig:Schematic} and
consists of three traps that are arranged in a linear array
\cite{1DRemark}. The two leftmost traps are simple harmonic
potentials, $V(x)=\frac{1}{2}m\omega^2x^2$, of identical trapping
frequency, $\omega$, and we will denote their ground states by
$|0\rangle_L$ and $|0\rangle_M$. The trap on the right is a double
well potential that (for numerical simplicity) we choose to be
composed of two simple harmonic potentials of frequency $\omega_R$,
with asymptotic ground states $|0\rangle_{R_L}$ and $|0\rangle_{R_R}$.
The distance between them is fixed and given by $d$.  The two lowest
lying eigenstates of the double well trap are the even and odd
combinations of these asymptotic ground states and are given by
$|0\rangle^\pm_R=(|0\rangle_{R_L}\pm|0\rangle_{R_R})/\sqrt{2}$.
Initially, a single atom of mass $m$ is located in $|0\rangle_L$ and
the other traps are empty.  After the STIRAP process has taken place,
the atom's wavefunction will be completely transferred into the double
well trap and possess a symmetry that is determined by the difference
between the trapping frequencies $\omega-\omega_R$.

The time dependence of the tunneling interaction is realised by
reducing and increasing the distance between the individual
traps \cite{Eckert:04}. We assume the middle trap to be fixed and the
outside traps to undergo the approach and reproach sequence. For the
distances between the respective traps, we assume the following time
dependence
\begin{align}
 d_{LM}(t)&=
\begin{cases}
d_{LM}& \text{if } t<\Delta t\\
\frac{1}{2}\left[\cos\left(\frac{2\pi t}{T}\right)+1\right]
           d_{LM}^\ast+d_{\text{min}}  & \text{if }t>\Delta t
\end{cases}
\label{eq:TimeSequence}
\end{align}
where $d_{LM}=d_{LM}(0)$ is the initial distance between the left hand
side and the middle trap, $d_\text{min}$ is the minimum distance
between the traps in the process and
$d_{LM}^\ast=d_{LM}(0)-d_\text{min}$. We always ensure that
$d_\text{min}>\max(\alpha,\alpha_R)$, where $\alpha$ and $\alpha_R$
are the ground state sizes of the harmonic potentials for $\omega$ and
$\omega_R$ respectively. The overall time for each individual approach
and reproach sequence is $T$ and the delay between the two sequences
is given by $\Delta t$. The sequence for $d_{MR}(t)$ is analogous to
eq.~\eqref{eq:TimeSequence} with the constant and the cosine part
reversed.

The main advantage of the counter-intuitive timing sequence is that it
relaxes the stringent conditions that would be necessary if one tried
to achieve transport by sequential two level tunneling. Any imprecise
timing would lead to the occurrence of Rabi type oscillations and,
thence, uncontrolled splitting of the wavefunction between all three
traps. In particular, the STIRAP method inherently avoids the transfer
of any part of the atom into $|0\rangle_M$ at any time during the
process. This is achieved by keeping the system adiabatically within a
dark state at all times, thereby avoiding any contribution from the
middle trap.

To explain the dynamics when the double well potential is involved, we
will make use of a so-called {\sl double} dark state that exists in
the Hamiltonian for such a system. If we assume that the energy of the
atom is fixed to the energy of the motional ground state of the left
trap at all times, $E=\frac{1}{2}\hbar\omega$, the system's
Hamiltonian can be written in terms of asymptotic eigenstates of the
individual harmonic traps as
\begin{equation}
  \label{eq:Hamiltonian}
  H=\hbar
  \begin{pmatrix}
    0          & -\Omega_{LM}(t) & 0 & 0 \\
    -\Omega_{LM}(t)  & 0 & -\Omega_{MR}(t) & 0\\
    0 & -\Omega_{MR}(t) &\omega-\omega_R & -\Omega_R \\
    0 & 0 & -\Omega_R &\omega-\omega_R
  \end{pmatrix}\;.
\end{equation}
Here $\Omega_{LM}(t)$ and $\Omega_{MR}(t)$ describe the time-dependent
tunneling frequencies between the states $|0\rangle_L$ and
$|0\rangle_M$ and $|0\rangle_M$ and $|0\rangle_{R}$ , respectively.
The tunneling frequency between the two states $|0\rangle_{R_L}$ and
$|0\rangle_{R_R}$ is given by $\Omega_R$ and is fixed at all times.
For numerical simplicity we model the traps as harmonic oscillator
potentials that have a fixed depth. In particular, we assume that this
depth does not change when the traps approach each other,
i.~e.~$V(x)=\min[V_L(x),V_M(x), V_R(x)]$.  Allowing for the potential
depth to change when the distance between the traps changes would
require a recalculation and readjustment of the trapping frequencies
at every moment in time, to ensure that the condition for the double
dark state (see below) is permanently fulfilled.  Recently it was
pointed out that a situation with fixed depth traps can be achieved in
optical traps by employing compensation lasers \cite{Eckert:06}.
  
The condition for the above Hamiltonian \eqref{eq:Hamiltonian} to have
an eigenstate with an eigenvalue equal to zero can be written as a
relation between the trapping frequencies and the tunneling frequency
within the double well trap \cite{Niu:04}
\begin{equation}
  \label{eq:ResonanceCondition}
  \omega-\omega_R=\pm\Omega_R\;.
\end{equation}
Since $\Omega_R$ is always positive, this condition implies that one
dark state exists for $\omega>\omega_R$ and one for $\omega<\omega_R$.
The respective eigenstates are given by
\begin{align}
  |\Phi^\pm\rangle&=\cos\theta|0\rangle_L
               -\sin\theta\left[(|0\rangle_{R_L}
               \pm|0\rangle_{R_R})/\sqrt 2\right]\\
              &=\cos\theta|0\rangle_L-\sin\theta|0\rangle_R^\pm\;,
\end{align}
where the mixing angle $\theta$ is defined as
\begin{equation}
  \tan\theta=\sqrt{2}\frac{\Omega_{LM}}{\Omega_{MR}}\;.
\end{equation}
Note that in the case of the trap on the right hand side being only a
single well trap, i.e. $\Omega_R=0$, this relation reduces to the
result of \cite{Eckert:04}, i.e.~all traps have to have identical
trapping frequencies.  If one fixes the distance between the two
individual traps in the double well trap to be $d_R=d\alpha_R$, where
$\alpha_R=\sqrt{\hbar/(m\omega_R)}$ is the width of the ground state
in the potential and $d>1$, the tunneling frequency within the double
well trap can be determined from the general relation for tunneling
between two harmonic traps \cite{Eckert:04}
\begin{equation}
  \frac{\Omega_R(d)}{\omega_R}=\frac{-1+e^{d^2}[1+d(1-\text{erf}(d)) ]}
                                      {\sqrt{\pi}(e^{2d^2}-1)/2d}\;.
\end{equation}
Combining this with condition \eqref{eq:ResonanceCondition}, the
resonance condition for the frequency, $\omega_R$, is given by
\begin{equation}
  \label{eq:omegaR}
  \omega_R=\frac{\omega}{1\pm\Omega_R}\;.
\end{equation}
This result can be easily interpreted. For the atom to move from the
left to the right, the system has to satisfy energy conservation.
While the asymptotic ground state energy levels for the double well
trap, $\frac{1}{2}\hbar\omega_R$, are either larger or smaller than
the ground state energies of the left and the middle trap,
$\frac{1}{2}\hbar\omega$, condition~\eqref{eq:ResonanceCondition}
ensures that either the eigenenergy of the symmetric or the
antisymmetric state of the double well is equal to
$\frac{1}{2}\hbar\omega$. In fact, for $\omega>\omega_R$ the particle
makes the transition into the (higher energy) antisymmetric state
($\frac{1}{2}\hbar\omega=E^-$), whereas for $\omega<\omega_R$ the
particle makes the transition into the (lower energy) symmetric state
($\frac{1}{2}\hbar\omega=E^+$)(see Fig.~\ref{fig:Schematic}).

\begin{figure}[tbp]
  \includegraphics[width=\linewidth,clip=true]{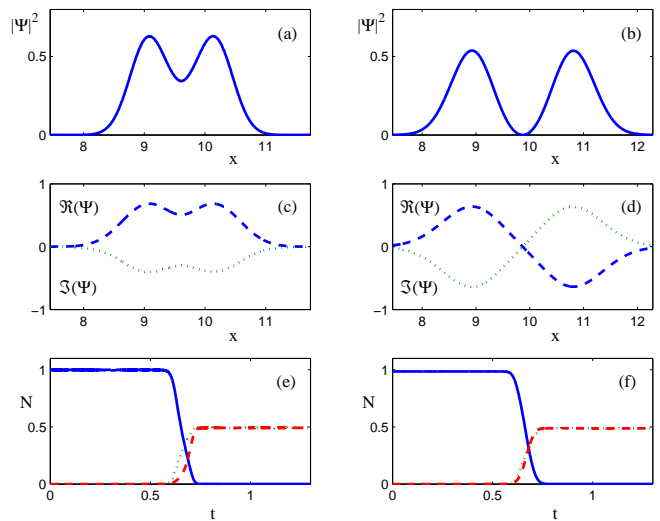}
  \caption{Wavefunctions at the end of the transfer process for the
    symmetric (left column) and the anti-symmetric (right column)
    case.  The panels (a) and (b) show the density and the panels (c)
    and (d) show the real (broken line) and the imaginary part (dotted
    line) of the wavefunction. The symmetries are clearly
    distinguishable. The bottom row shows the change of population in
    the left trap (full line) and the two halves of the double well
    trap (broken line and dotted line). The spatial co-ordinate is
    scaled in units of $\alpha$.}
\label{fig:FinalWF}
\end{figure}

To demonstrate the creation of coherent superposition states we have
performed numerical integrations of the Schr\"odinger equation of the
system, using an FFT/split-operator algorithm. At the start of each
simulation the initial state is given by a Gaussian wavefunction
centered in the trap on the left hand side and the distance between
the two traps in the double well is given by $d_R=1.5 \alpha_R$.
Typical results are shown in Fig.~\ref{fig:FinalWF}, where the time
delay between the approach sequences is given by 10\% of the time for
each approach process, which was in turn chosen to be
$T=(1.3\Omega_R)^{-1}$. The time dependence of the distance between
the traps is chosen according to eq.~\eqref{eq:TimeSequence} and the
minimum distance between the traps is fixed at
1.2\;$\max(\alpha,\alpha_R)$, to make sure tunneling is the only
method for transfer. The three panels on the left correspond to the
case $\omega_R>\omega$ and the three panels on the right to the case
$\omega_R<\omega$. The even and odd symmetries, respectively, of the
final wavefunction are clearly visible. The two panels in the bottom
row show the amount of $|\psi|^2$ in each trap. While at the beginning
the wavefunction is completely localised in the trap on the left, at
the end of the process it is equally split between the two wells of
the double trap. There is never any significant population in the
middle trap (not shown). We have performed these simulations for a wide
range of values for $d_R$ and the delay interval, $\Delta t$, and
found this process to be extremely robust.

\begin{figure}[tbp]
  \includegraphics[width=\linewidth,clip=true]{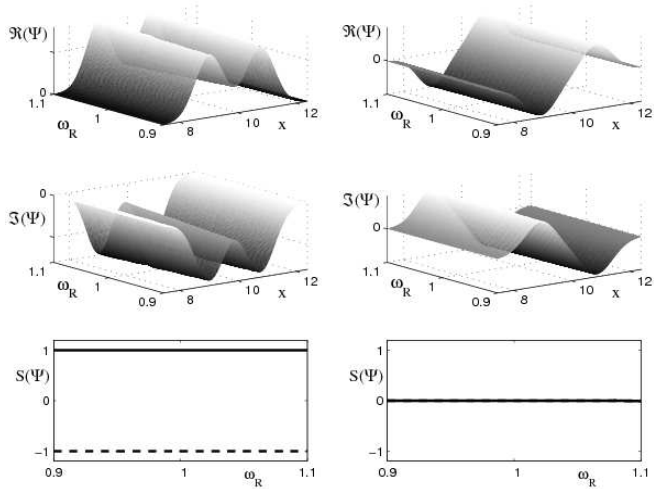}
  \caption{The four upper panels show the real and imaginary parts of
    the final wavefunction as a function of the trapping frequency in
    the double well trap. The trap frequency $\omega_R$ is normalised
    to the value required by the resonance condition and it is varied
    in an interval of $\pm$10\%. The two graphs on the left correspond
    to the case $\omega<\omega_R$ and the two on the right to
    $\omega>\omega_R$. The two panels in the bottom row show the
    symmetry function for the wavefunctions above as defined in
    eq.~\eqref{eq:SymASym}, where the full line represents the real
    parts and the broken line the imaginary parts.  Both lines are
    lying on top of each other for the antisymmetric case.}
  \label{fig:Noise} 
\end{figure} 

To examine the systems fragility to noise, we simulate the process
above using a frequency range for $\omega_R$ that is $\pm$ 10\% out of
resonance with the value that the symmetric or the antisymmetric
resonance condition \eqref{eq:ResonanceCondition} demands. In
Fig.~\ref{fig:Noise} the upper two panels on the left (right) show the
real and imaginary parts of the final wavefunction for
$\omega_R>\omega$ ($\omega_R<\omega$). While the transferred amount is
no longer 100\% (however, it is still $>99$\%) for the non-resonant
systems, one can immediately see that the symmetry is still a
preserved property. To quantify the symmetry we define the following
functions
\begin{equation}
  \label{eq:SymASym} 
  S_R=\frac{\sum_z\Re(\psi(z))}{\sum_z|\Re(\psi(z))|}\qquad,\qquad
  S_I=\frac{\sum_z\Im(\psi(z))}{\sum_z|\Im(\psi(z))|}\;,
\end{equation} 
which give $S_{R,I}=\pm 1$ (depending on the phase of the
wavefunction) for a perfectly symmetric state and $S_{R,I}=0$ for a
perfectly antisymmetric state. The lowest panel on the left (right)
hand side shows this function for $\omega_R>\omega$
($\omega_R<\omega$) and confirms the optical inspection of the upper
panels that the symmetry of the wavefunction is very robust against
imperfections in the setup.

\begin{figure}[tbp]
  \includegraphics[width=0.8\linewidth,clip=true]{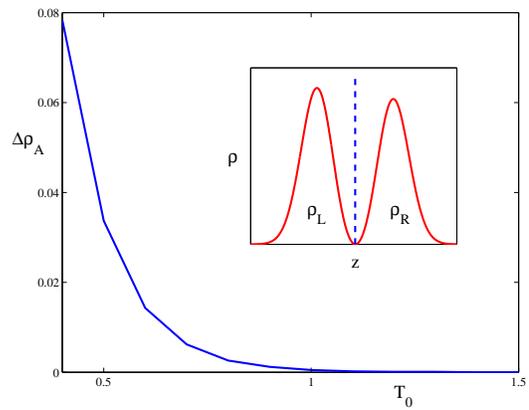}
  \caption{Oscillating density imbalance between the two halves of the
    double well trap, $\Delta\rho_A=\max(\rho_L-\rho_R)$. The
    curve is for the anti-symmetric situation and plotted against the
    time interval of the process in units of $\frac{1}{\Omega_R}$.
    The inset shows an imbalanced wavefunction for which $\rho_L$
    and $\rho_R$ describe the density fractions towards the left
    and the right from the center of the trap respectively.}
  \label{fig:Imbalance} 
\end{figure}

The condition that the transfer has to be performed adiabatically is
already inherent in the name STIRAP. The whole process therefore has
to proceed slower than the inverse of the lowest trapping frequency to
avoid excitations of higher lying levels
\begin{equation}
  T+\Delta t>\frac{1}{\min[\omega,\omega_R]}\;.
\end{equation}
For our setup, which includes the double trap, this condition only
ensures that a full transfer will be achieved from the left trap into
the double well trap. In order to achieve a steady state with the
desired symmetry, a second condition on the timescale has to be
fulfilled, which ensures that the process is slow with respect to the
tunneling frequency between the two wells in the double well trap
\begin{equation}
  T+\Delta t>\frac{1}{\Omega_R}\;.
\label{eq:ad2}
\end{equation}
If this condition is not fulfilled, the final state will no longer be
an eigenstate but rather be given by an oscillating population
imbalance between the two wells. This, however, does not affect the
symmetry of the final state. In Fig.~\ref{fig:Imbalance} we show the
amplitude of the oscillations
$\Delta\rho_A=\max(\rho_L-\rho_R)$ for different values of
$T_0=T+\Delta t$. It can be clearly seen that a steady state is
established once condition \eqref{eq:ad2} is fulfilled. While the data
shown are for the case $\omega>\omega_R$, the behaviour is analogous
for $\omega<\omega_R$.

The above process shows that spatial STIRAP-like processes do not only
allows for a robust transport or preparation of the wavefunctions
amplitude, but also allow for high fidelity when controlling the
phase.  It is therefore a good starting point for many applications in
atom interferometry \cite{Zhang:06}. The possibility of manipulating
the parts of the wavefunction selectively with well-focused laser
beams and closing the interferometer by running the whole process in
reverse then opens the possibility of creating universal quantum gates
\cite{Englert:01}.

Schemes like this pose a challenge to current technologies, since they
require dynamical control over the position of the micro-traps.
Several different technologies have recently emerged that allow for
such control and active experimental efforts are undertaken in many
laboratories. Initial experiments have already shown the possibility
of dynamically controlling the distance between traps \cite{Boyer:06}.

In summary we have suggested a robust and straightforward technique
for the controlled creation of center-of-mass superposition states of
atoms in micro-traps. By adjusting the trapping frequencies the final
wavefunction can be chosen to be symmetric or antisymmetric. This
process relies on a STIRAP-like method for transferring the atom into
the new state, and the appropriate adjustment of trapping frequencies
in order to choose the symmetry. While the spatial STIRAP process has
been shown to be very robust for transfer of amplitudes, we have shown
that it can also be used to create a very robust technique to control
the phase of the wavefunction.

\begin{acknowledgments}
  The work was supported by the China-Ireland Research Collaboration
  Fund of Science Foundation Ireland and the Ministry of Science and
  Technology of the People's Republic of China and by Science
  Foundation Ireland under project number 05/IN/I852. KD acknowledges
  generous support from CIT.
\end{acknowledgments}



\begin{thebibliography}{99}
\bibitem{1DRemark} Since the dynamics in our process can be arranged
  to lie in a single plane, the restriction to one-dimension for
  computational purposes does not have any implications for the
  generality of our method.
\end{thebibliography}
\end{document}